# CERN, a working example of global scientific collaboration

Jurgen Schukraft, *CERN, 1211 Geneva 23, Switzerland* (schukraft@cern.ch)

*Abstract*—The topic of this conference is 'South-South and North-South Collaboration in Science and Technology', which is addressed in this contribution in the context of basic research in high energy physics (HEP). The question whether developing countries can or should invest scarce resources in big science is not covered. HEP may be less expensive than one might fear, but cheap it is not, so priorities have to be set and these may indeed differ from country to country. The scope of this article is not to argue one way or another, but rather to give an indication and practical examples of both the *requirements* and the *opportunities* for scientific collaboration with CERN.

I. CERN

The European Organization for Nuclear Research (CERN) is an intergovernmental organization with 20 Member States (Austria, Belgium, Bulgaria, Czech Republic, Denmark, Finland, France, Germany, Greece, Hungary, Italy, The Netherlands, Norway, Poland, Portugal, Slovak Republic, Spain, Sweden, Switzerland and the United Kingdom), located in Geneva, Switzerland, but straddling the French-Swiss border. It is the world's largest laboratory for fundamental research in particle physics with a support staff of ~2500 and a yearly budget of ~ 1 billion CHF. Its mission is to provide facilities and infrastructure for carrying out experiments in basic research, primarily high energy physics (HEP) but to a smaller extent also nuclear physics, astro-particle physics and theory, to answer fundamental questions concerning structure and composition of **matter** ('*What is the Universe made of*') and **forces** ('*What keeps it together*').

CERN was founded in 1954 with the help of UNESCO with the prime aim to pool European resources for basic *research* which would be beyond the means of individual countries. Less explicit, but at least of equal importance, were *economical* and *political* considerations in the aftermath of Word War II: to create a center of excellence which would make and keep Europe competitive in leading edge science and technology, stop the brain-drain of its most educated and creative citizens and rebuilt trust, collaboration and coherence between the nations of Europe.

Celebrating this year its 50$^{th}$ anniversary, CERN is considered a success almost beyond the most optimistic expectations: In *research*, it is today the foremost center for high energy physics in the world with several Nobel Prize winning results; it has not only stopped but reversed the brain drain (more than half of the world's particle physicists participate in its programs). In terms of *economical* benefits, it is a vibrant incubator for R&D and technology transfer who's impact is not only limited to Europe (the *Word Wide Web* was invented at CERN in the 90's as a communication tool for scientists collaborating across the world). On the *political* level, CERN is a most successful and effective example of international collaboration which served as a blueprint for other research organizations (ESA, EMBL, ESRF). During the cold war, a scientific collaboration between West and East could be established via CERN which kept open one of the very few communication channels between citizens across the iron curtain.

Given its excellent record in the past and present, what does the future hold for CERN? On the research frontier, CERN is currently building the most powerful accelerator ever, the Large Hadron Collider LHC. It will come into operation in 2007 and run for at least 10 – 15 years, keeping research at CERN at the cutting edge well into the next decade. CERN has since long outgrown the *N* (for Nuclear) in its name by going to ever higher energies, and now it has effectively left behind the *E* (for European) with the LHC machine and its four large experiments (named ALICE, ATLAS, CMS and LHCb). Both the machine and even more so the detectors are designed, constructed and exploited by a global, world-wide Collaboration of scientists and engineers, which by now encompasses about 60 countries, including very significant participation from the 'South' and 'East'. Some 2000 physicists from non-member states participate today in the CERN program, corresponding to about 1/3 of all CERN users. While CERN is not likely to change its name or its administrative structure as a European organization, effectively it has reached out to the world, both developed and developing countries alike. This brings with it new opportunities, but new challenges as well.

II. CERN USERS: EXPERIMENTS & COLLABORATIONS

CERN, as a *facility*, provides the large scientific tools and infrastructure needed to carry out experiments in high energy physics by its *users*. These users are based in their respective home countries in university or research labs (CERN employs only a very small number of researchers, and most of them for a limited duration only). Research is both initiated and executed by these users, after review and ranking by scientific peer committees. The users organize themselves around individual experiments and collaborations, which are groups of scientists ranging in size from below 10 to almost 2000 people. These collaborations propose, design, build and operate their experiment and all members of the collaboration have free and unfettered access to its data and results, which in the end are made public in the scientific literature. Funding for both construction and operation of the experiment is provided by national funding agencies, in general directly to the individual participating groups. Otherwise the Collaborations are administratively and scientifically rather independent and self-governing, i.e. they decide who can take part, who does what.

The four experiments at the LHC are amongst the most complex and challenging technical endeavors ever undertaken, requiring many years of R&D to push technology well beyond existing limits, and many years of construction by literally thousand of experts (the smallest experiment includes more than 500 people, the largest almost 2000 !). Almost 20 years will have passed from first ideas to final realization when the experiments will actually start taking data in 2007. Technologies employed are all state-of-the-art (and sometimes just beyond) in a number of areas: From *mechanics and engineering* (micrometer precision in objects of tens of meters, lightweight but stable structures), *advanced materials* in a hostile environment (radiation and chemical), *sensor technology* (from micrometer sized silicon sensors to hundreds of tons of heavy metals), *micro-electronics* (custom designed VLSI circuits and electronic boards) to *information technology* (virtual communication & document sharing, large scale distributed computing with the *GRID*). All these technologies have obvious potential for technology transfer and industrial spin-off.

A practical example of how such a collaboration works, and in particular how the 'nations of the south' can and do participate in this frontier research, is illustrated by the 'midsize' ALICE experiment. It consists of about 1000 members (2/3 from CERN member states, 1/3 from outside) from over 80 institutions in ~30 different countries. The construction cost is ~ 150 M CHF. While Italy is the largest country in ALICE (with 160 scientists from 12 institutions), contributing well over 20% of the construction funding, very significant participation and contributions are made by 'countries of the south', ranging from India (55 people, 7 institutes, 3 M CHF) over China, Mexico, and South Africa to Cuba (1 Institute, 4 scientists, 0.045 M CHF). India, which has been a member of ALICE since 1992 and contributes both scientifically and financially at the level of a number of CERN member states in ALICE, has slowly but very steadily built up an international scientific reputation in the field explored by the experiment; it contributes with original research and state-of-the-art equipment (from sophisticated sensor technology to VLSI electronics) developed and manufactured using indigenous talent and industry. On the other hand, South Africa and Cuba have been members since 2003 only; they have started on a more modest scale (with 1-2 staff and many students) and are involved, for the time being, in less capital intense activities like computing and physics.

### III. REQUIREMENTS AND OPPORTUNITIES

The ALICE example shows that participation in research even on the scale of an LHC experiment can come over a large range of topics and scales. Experiments expect from their collaborators that they are motivated by *scientific interest* (exceptionally by technology for 'associated institutes'), that they are capable of supporting their activity largely by their *own means* and funding, and that they are *actively* participating by contribution *talent and resources* within their capabilities. Sheer size matters less, and a new activity usually starts on a small scale, sometimes as small as one single group with few staff and a number of students. A minimum infrastructure at the home institute has to be in place (or has to be created), most importantly for good communication (computers and networking). Indigenous hardware contributions will in general need access to workshops and/or local industry. Funding is required both locally (salaries, infrastructure, international travel) and for investment in the experiment. The latter will differ from case to case but can start as low as 10,000 CHF per year in justified cases. A strong backing and support by the relevant national authorities and funding agencies is very important. International collaboration on this scale is governed by a Memorandum of Understanding which is in general signed at the national level; while not legally binding it implies both benefits and commitments over a long time scale.

The most important motivation (and benefit) for participating in an LHC experiment for the individual researchers is (or should be) scientific: To be a *participant* rather than a *spectator* in one of the most exciting and fundamental endeavors undertaken by the inquisitive human mind. This may however not be sufficient reason to get the necessary resources. While it is not within the scope of this contribution to dwell in any depth on the various other benefits and opportunities for developing nations of participating in frontier HEP experiments, a few shall be listed briefly. High amongst them are the ones mentioned previously as one of the reasons for creating CERN in the first place: Even if the actual experiments are carried out outside national borders, participation will provide ample research activity and infrastructure based *at home*, which is crucial to develop and keep a resident scientific elite, allow the 'good' to become 'excellent' (and internationally known), stop the brain drain and maybe attract back some émigrés. Training and education of students is one of the main missions of CERN with over 1000 coming every year for typical periods of several months up to 3 years. They receive 'hands on' training in a variety of technologies in an international environment, with clear objectives and deliverables, strict selection criteria and in general return, at the end, to the institute of origin. The large LHC collaborations are also an excellent place for 'international networking', as they allow (and in fact require) a multitude of personal and institutional contact and collaboration all over the world.

Technology transfer, spin-off and industrial return are subjects which have been extensively discussed and studied; they are recognized as a major benefit for most companies dealing with CERN and its experiments. Of relevance, in particular for developing countries, is the direct involvement of nationals and their incentive to involve local industry, for either convenience or price reasons. The often unique and 'custom tailored' specifications open a window of opportunity for small or midsized specialized industry, even if not internationally known. To quote again some specific examples, ALICE has, with the help of national physicists, found excellent and very competitive providers of tungsten in

China, crystals in Russia, power cables in India, micro-cables in the Ukraine, …

IV.   Conclusion

High Energy Physics (HEP), as carried out at CERN, is science in the 'top league', at the cutting edge of science and technology. Participation requires talent, resources, and a long term commitment. HEP is BIG science, in terms of scale but also in terms of importance and scope. Nevertheless, participation on a small scale is possible and even welcome. HEP is international and collaborative science and being part of one experiment means being connected to much of the world. CERN and its experiments are open to the scientists of the world, offering access to a unique research infrastructure and participation in multimillion dollar experiments at the absolute forefront of science and technology. HEP is a truly global venture, in both spirit and in execution; if governments are willing to support it, scientists can make it work.